\documentclass[twocolumn,amsmath,amssymb,prb,superscriptaddress,a4paper,floa
tfix,showkeys]{revtex4}
\usepackage{epsfig}
\usepackage{graphicx}
\usepackage{amsmath,amssymb}
\usepackage{subfigure}
\usepackage{graphicx,color}


\begin{document}
\bibliographystyle{unsrt}
 \title
{An overview on Single Apparatus Quantum Measurements}
\author{B. Mehmani$^{\,1)}$, and Th. M. Nieuwenhuizen$^{\,1)}$ }
\affiliation{$^{1)}$ Institute for Theoretical Physics,
Valckenierstraat 65, 1018 XE Amsterdam, The Netherlands}
\begin{abstract}
Given the state of a quantum system, one can calculate the expectation value of any observable of the system. However, the inverse problem of determining the state by performing different measurements is not a trivial task. In various experimental setups it is reasonably straightforward to reconstruct the state of a quantum system employing linear tomographic technique. In this way the elements of the density matrix can be linearly related to a set of measured quantities.
But since different observables of a quantum system are not commuting with each other, one often has to perform series of successive measurements of observables which cannot be done simultaneously. Simultaneous measurement of observables cost less time and energy and is more beneficial. In this paper we review the strategy of quantum state tomography with simultaneous measurement of commuting observables. This can be done by introducing an assistant system of which the state is known. We show that the interaction between the assistant and the system of interest within different frame works allows the reconstruction of the state of the system. Specifically, we consider a two-level system and reconstruct its initial state by introducing  an  assistant which can be either another two-level system or a single cavity mode of the electromagnetic field. 
\end{abstract}
\maketitle

\section{\label{sec:state}Introduction}
In classical physics, the state of a system is characterized by
specifying the values of all physical quantities, for instance
the positions and the velocities of the particles that constitute
the system. In quantum mechanics the situation is complicated by
the fact that the physical quantities are mathematically
represented by specific type of operators called observables, which in general are elements of a
\textit{non-commutative algebra}. Hence their values cannot be
simultaneously specified, as emphasized in the Heisenberg's
uncertainty principle. Instead, the measurement results of each observable is characterized by a
probability distribution, which involves statistical fluctuations.
The ``state of the system'' is then represented by a mathematical
notion that allows us to express the probability distribution of
all the observables for an ensemble of identically prepared systems.

This is best described in the statistical interpretation of quantum
mechanics~\cite{BALLENTINE1970,Muynck2002,Balian1989}, to which we adhere. In this interpretation, the state of a
system is a mathematical object from which we can derive any
probabilistic prediction about the physical quantities attached to
this system. One typically imagines some experimental apparatus and procedure which ``prepares'' this quantum state; the mathematical object then reflects the setup of the apparatus. This way, the quantum state
 accounts for the full information
available about the preparation of the system, from which we wish
to derive consequences for future experiments. Since this knowledge
is probabilistic it does not refer to a single system or a single
event. Thus what we call a state, which is most of the time a mixed one,
characterizes a statistical ensemble of systems of the same type,
which are all prepared under identical physical considerations. The state is thus a mathematical representation of the result
of a certain state preparation procedure; it accounts for our
information about this preparation and upon knowing it we can
elaborate consistent probabilistic predictions. A
standard tool to implement the statistical definition of state is the density matrix,
which generalizes the pure state represented by a wave function.
Indeed, there is no conceptual difference between wave function and density matrix
which are both mathematical means for evaluating expectation
values of the observables of the system or probabilities.

In the frame work of the statistical interpretation, the laws of
quantum mechanics can be summarized as follows~\cite{BALLENTINE1970,Muynck2002,Balian1989}:
\begin{itemize}
\item
An \index{Observable}observable $\hat{{\cal O}}$
  is represented by a self-adjoint linear operator acting on the Hilbert space pertaining to the system.
It has a spectral representation, $\hat{{\cal O}} =
\sum_{i}o_i\hat{P}_i$ where $o_i$ are the eigenvalues
of $\hat{{\cal O}}$ and $\hat{P}_i$ are the orthogonal projection
operators related to the orthonormal eigenvectors of
$\hat{{\cal O}}$, i.e., $\hat{P}_i = \sum_{m}\vert m, o_i\rangle \langle m,
o_i\vert$. The parameter $m$ labels the degenerate
eigenvectors of $\hat{{\cal O}}$.
\item
The state of a system at a given time is represented by its
\index{Density matrix}density matrix, $\hat{\rho}$, which is a self-adjoint operator in
Hilbert space with a unit trace. The density matrix should also be
semipositive to ensure that any variance of the observables of the
system is non-negative. Pure states correspond to the special case
\begin{align}
 \hat{\rho}^2 = \hat{\rho}.
\end{align}

\item
The dynamics of the system can be obtained by
\begin{align}
 \hat{\rho}(t) =
\hat{U}(t,t_0)\,\hat{\rho}(t_0)\,\hat{U}^{\dagger}(t,t_0),
\end{align}
where $\hat{U}(t,t_0)$ is
the \index{Unitary time-evolution operator}unitary time evolution operator.
\item
Given the density matrix $\hat{\rho}$ of a system, one can find the
\index{Expectation value}expectation value of any observable $\hat{{\cal O}}$ of the system
in the considered situation as
\begin{align}
 \langle \hat{{\cal O}}\rangle = \textnormal{tr}
[ \hat{\rho}\, \hat{{\cal O}} ],
\end{align}
 where $\textnormal{tr}[\cdots]$ stands for the trace of a matrix.
\end{itemize}
Let us emphasize that through out this paper the operators are always distinguished by a $\hat{}$ sign.\newline
Now we wish to face an inverse problem. Consider an
ensemble of systems \textrm{S} prepared in some well-defined, but
unknown, fashion. Nothing is a priori known about their state
$\hat{\rho}$, in other words, the probability to observe some result or
another in the measurement of an observable is unknown. The following
question then is of our interest. How can one determine the density matrix of
this ensemble by identification of a set of commuting
observables, the measurement of which permits the precise
determination of $\hat{\rho}$? In other words, how can one determine the quantum
statistical operator that describes the preparation of the system?

Procedures of reconstructing the quantum state from
measurements are known as \index{Quantum state tomography}{\it quantum state tomography}.
Recently, they have found some applications in quantum information processing~\cite{Nielsen2000}. For example, in quantum cryptography
one needs a complete specification of the qubit
state both as it is emitted from the source and as it is received after
transmission~\cite{Pasquinucci2000}.

In the simplest example
of a \index{Two-level system}spin-$\frac{1}{2}$ system
or equivalently any two-level quantum system the state is described by a $2 \times 2$
matrix.
In the two-dimensional Hilbert space, any observable is a linear
combination of Pauli operators, which satisfy
\begin{eqnarray}\label{ch2,eq:pauli_algebra}
 &&\hat{\sigma}_{\alpha}^{2} = \hat{1}, \hspace{10pt} \alpha = x,
y, z,\nonumber\\
&&\left[ \hat{\sigma}_x, \hat{\sigma}_y \right]= i\hat{\sigma}_z,
\end{eqnarray}
 and are represented by the Pauli matrices
\begin{eqnarray}\label{ch2,eq:Pauli_matrices}
 &&\hat{\sigma}_x =\begin{pmatrix}
  0 & 1 \\
  1 & 0
\end{pmatrix},\qquad
\hat{\sigma}_y = \begin{pmatrix}
 0 & -i \\
  i & 0
\end{pmatrix},\qquad
\hat{\sigma}_z = \begin{pmatrix}
  1 & 0 \\
  0 & -1
\end{pmatrix}.\nonumber\\
 \end{eqnarray}

A state is characterized by three real numbers: one for the
  diagonal elements of the $2\times2$ density matrix $\hat{\rho}$, and two
  for its off-diagonal elements. Equivalently, we can introduce
  the polarization vector, $\vec r$, the components of which
  are the expectation values of the Pauli matrices.
  \begin{equation}\label{ch2,eq:polarization_vector}
r_\alpha = \textnormal{tr}\left[\hat{\rho}\, \hat{\sigma}_{\alpha}\right], \,\,\,\,\,\,\,\,\,\,\alpha=x, y, z.
 \end{equation}

Once we know the value of these parameters, we are able to determine the value of the density matrix, making use of the identity

\begin{equation}\label{ch2,eq:density_matrix_def}
 \hat{\rho} = \frac{1}{2} \left({\mathbf 1} + \vec r \cdot \vec{\hat{\sigma}} \right).
\end{equation}

Thus, according to the above argument, one has to perform three
incompatible measurements for the unknown state determination, e.g.,
measuring the spin components along the \textrm{x}-, \textrm{y}- and
\textrm{z}- axes via Stern-Gerlach setup. However, during the
measurement procedure of each component one looses the information about
the two other components, since the spin operators in different
directions do not commute. Thus, to determine the state of a
spin-$\frac{1}{2}$ system, one needs to use three sets of Stern--Gerlach measurements
performed along orthogonal directions. In this approach, the state of any two-level
system, represented by a $2 \times 2$ density matrix $\hat{\rho}$, can be
fully determined only through measurement of three linearly
independent observables
which do not commute and cannot be simultaneously measured.

It has been proven  that the unknown density matrix
of such a system \textrm{S}, in particular the full polarization
vector of a spin-$\frac{1}{2}$ system, can be determined indirectly. This can be done by means of a
single set of measurements performed simultaneously on \textrm{S}
and an auxiliary system (assistant or ancilla). \index{Assistant}The assistant (or ancilla) starts its evolution from a known state~\cite{Allahverdyan2004,D'Ariano2002,Mehmani2008,Peng2007}

The suggested strategy is the following: Initially \textrm{S} lies in the
unknown state that we wish to determine, while the state of the assistant
\textrm{A} is known. During some time lapse
\textrm{S} and \textrm{A} interact in a known fashion. As a result their joint
state is modified: it involves correlations and keeps memory of
the initial state of \textrm{S}. Two
commuting observables of the combined system (system + assistant) are
then simultaneously measured. Repeating this process provides then the necessary statistical
data: the expectation values of the observables and also
their correlation. We will show that one can infer the three
components of the initial polarization vector of \textrm{S}, and hence the state of the system from
these three sets of data.
 This type of information transfer is  remarkable because initially an unknown information was embedded in the matrix
elements of $\hat{\rho}$, or equivalently in the components of the
polarization vector of \textrm{S}. It had a \textit{quantum
nature}, and could not be represented by an ordinary probability
distribution, due to the non-commutation of the three components
of the spin operator. After the interaction between \textrm{S} and
\textrm{A} this unknown initial information about \textrm{S},
together with the known information about \textrm{A}, is
redistributed among the matrix elements of the joint density
matrix of the overall system, \textrm{S+A}. However, the resulting \textit{classical}
joint probability distribution for the observables of
the system \textrm{S} and the assistant \textrm{A} can keep full memory of the
initial quantum information about \textrm{S}. The process on which
we rely, amounts to a transformation of quantum
information into classical information, which can be gained by a
classical type of measurement involving commuting observables
only. This measurement modifies the state of \textrm{S+A}, but it can
preserve all the matrix elements of $\hat{\rho}$.
The idea of transforming quantum into classical information by
using an assistant system \textrm{A} was first proposed by D'
Ariano~\cite{D'Ariano2002} who showed the possibility of mapping the
density matrix of \textrm{S} onto a single observable of
\textrm{S+A}. It was explicitly implemented in a dynamical form by Allahverdyan {\it et al}.
~\cite{Allahverdyan2004}.
In particular, the authors~\cite{Allahverdyan2004} showed that one
can determine the unknown state of a spin-$\frac{1}{2}$ system with a single
apparatus by using another spin-$\frac{1}{2}$ assistant. This idea was recently implemented by Peng {\it et al.}~\cite{Peng2007} who used pulses to induce the proper dynamics of the interaction between the spin-$\frac{1}{2}$ system and its assistant.  They verified the initial state of the system obtained from this procedure with the result of direct measurements of the three components of the spin vector of the system.\newline
Aquino {\it et al.}~\cite{Aquino2007} and Mehmani {\it et al.}~\cite{Mehmani2008} showed that the initial state of a two-level system can be characterized by introducing a single mode of a coherent light instead of introducing another spin, so the dimension of the Hilbert space of the assistant can be larger than that of the system. More specifically, Aquino {\it et al.} showed that in the resonant case, i.e., when the frequency of the two-level system is equal to the cavity mode frequency, it is not possible to determine the initial state of a two-level system by measuring the the energies of the system and the assistant. However, this scheme is still applicable for spin-$\frac{1}{2}$ systems because in order to recover the initial state of the spin-$\frac{1}{2}$ system one can measure the transversal component of spin ($x$ or $y$ component) and the number of the photons in the cavity~\cite{Aquino2007}. In general, for two-level systems other than spins, it is rather difficult to measure the transversal component since it cannot be defined well. While the $z$-component of quasi-spin is related to the level occupation and thus the energy of the two-level system.
The crucial point in this case is that there should be a detuning between the frequency of the field and  that of the system of interest in order to invert the relavent relations between simultaneously measured observables in one hand and the elements of the initial density matrix of the two-level system on the other hand~\cite{Mehmani2008}. 

This overview is organized as the following:
In section~\ref{spin_assistant} we review the idea of determining the state of a spin-$\frac{1}{2}$ system by letting it interact with another spin-$\frac{1}{2}$ system which starts its evolution from either a completely disordered state or a pure state.
Section~\ref{light_assistant} is devoted to the situation where the assistant is considered to be a coherent single mode of the electromagnetic field. The last section~\ref{con} presents our conclusion. 

\section{\label{spin_assistant}Spin as an Assistant}
As it was mentioned before, the
aim is to find an indirect procedure to determine the unknown state of a quantum system. The desired procedure only consists of measurements
of commuting observables, which therefore can be performed by
means of a single apparatus. For the quantum system we consider a two-level system, \textrm{S}, the state of which we wish to determine.
Let the system \textrm{S} be coupled to an auxiliary two-level 
system \textrm{A}. \textrm{A} is in a known state. At later time we  measure the value of one observable of $S$ and one of $A$. Since the two observables belong to two different Hilbert spaces, they commute and therefor the measurement process can be performed simultaneously. Moreover, the correlator of two observables can be read off from coincidences.

Let the initial unknown state of \textrm{S} be a density matrix of the form
\begin{equation}
 \label{S_initial_state}
\hat{\rho} = \frac{1}{2} \left(1 + \vec{r} \cdot \vec{\hat{\sigma}}\right),
\end{equation}
where
\begin{equation}
 \label{ave_def}
r_i \stackrel{def}{=} \textnormal{tr} \left[\hat{\rho}\, \hat{\sigma}_i\right],\qquad i = x, y, z.
\end{equation}
The state is called pure if $\vert \vec r \vert = 1$, in which case the eigenvalues of $\hat \rho$ are $0$ and $1$. $\vert \vec r \vert < 1$ represents a mixed state, and $\vert \vec r \vert >1 $ is physically excluded.

We choose the state of the assistant, represented by $\hat{R}$ as
\begin{equation}
 \label{A_initial_state}
\hat{R} = \frac{1}{2} \left(1 + \lambda \hat{s}_z\right),\qquad 0 \leq \lambda \leq 1
\end{equation}
where $\hat s_x$, $\hat s_y$, and $\hat s_z$ are the Pauli matrices in the Hilbert space belonging to the assistant $\textrm{A}$.\newline
Initially there is no interaction between $\textrm{S}$ and $\textrm{A}$. Therefore the initial state of the overall system, $\hat{\Omega}_0$, can be written as
\begin{equation}
 \label{S+A_initial_state}
\hat{\Omega}_0 = \hat{R} \,\hat{\rho} = \frac{1}{4}\begin{pmatrix}
                                          (1 + \lambda) (1 + \vec{r} \cdot \vec{\hat{\sigma}})& 0\\
						0 &  (1 - \lambda) (1 + \vec{r} \cdot \vec{\hat{\sigma}})
                                         \end{pmatrix}.
\end{equation}
Now we let the two systems interact for some time. The interaction can be described with the help of a $4 \times 4$ unitary matrix $\hat{U} = e^{- i \hat{H} }$, where we set $t = 1$. Here, we don't specify our Hamiltonian and consider a general unitary matrix and we parametrize it such that it generates a proper time-evolved overall density matrix at later time $t = 1$, given by $\hat{\Omega}_f$ such that the initial state of $\textrm{S}$ can be read off easily. The observables of which the measurements yields the determination of the initial state of $\textrm{S}$ are the final polarization of each spin of the overall system $\textrm{S+A}$~\cite{Allahverdyan2004}. They can be measured simultaneously and the correlator of the two can be derived from the gathered data. We show that it is possible to read off the initially unknown state of the system \textrm{S} from the three above mentioned sets of data.
Let us decompose $\hat{U}$ into the following $2 \times 2$ block matrix,
 \begin{equation}
  \label{general_U}
\hat{U} = \begin{pmatrix}
 \hat{A} & \hat{C} \\
  \hat{B} & \hat{D}
\end{pmatrix},
 \end{equation}
  and express the unitarity of $\hat U$ in terms of the
$2\times2$ matrices $\hat A$, $\hat B$, $\hat C$, $\hat D$ in the Hilbert space of
\textrm{S}.
The polar decomposition of $\hat{A}$ and $\hat{B}$ yields
\begin{equation}
 \label{A,B_polar_dec}
\hat{A} = \hat v \hat k, \qquad \hat B = \hat w \hat k',
\end{equation}
where $\hat v$, and $\hat w$ are unitary $2 \times 2$ matrices while $\hat k$ and $\hat k'$ are semi-positive Hermitian $2 \times 2$ matrices. Since $\hat v$, and $\hat w$ are unitary, it is easy to see that $\hat k$ and $\hat k'$ are the non-negative square roots of $\hat A^\dagger \hat A$ and $\hat B^\dagger \hat B$, respectively. If $\hat k$ and
$\hat k'$ have a vanishing eigenvalue, these representations of $\hat A$
and $\hat B$ still hold but are no longer unique. We shall restrict ourselves to the case where $\hat k$ and
$\hat k'$ are strictly positive. \newline
The condition $\hat U \hat U^\dagger = 1$ implies
\begin{eqnarray}
 \label{unitary_U_1}
&&\hat C \hat C^\dagger = 1 - \hat A \hat A^\dagger ,\qquad \hat D \hat D^\dagger = 1 - \hat B \hat B^\dagger,\\
&&\hat A \hat B^\dagger + \hat C \hat D^\dagger = 0,
\end{eqnarray}
while $\hat U^\dagger \hat U = 1$ implies
\begin{eqnarray}
 \label{unitary_U_2}
&&\hat A^\dagger \hat A + \hat B^\dagger \hat B = 1, \qquad \hat C^\dagger \hat C + \hat D^\dagger \hat D = 1,\\
&& \hat A^\dagger \hat C + \hat B^\dagger \hat D = 0.
\end{eqnarray}
Implementing (\ref{unitary_U_2}) on the polar decomposition of $\hat A$ and $\hat B$ given by (\ref{A,B_polar_dec}) yields
\begin{equation}
 \label{k,k'}
\hat k' = \sqrt{1 - \hat k^2}.
\end{equation}
Thus $\hat C \hat C^\dagger$ and $\hat D \hat D^\dagger$ can be simplified as
\begin{equation}
 \label{C,D,dec_1}
\hat C \hat C^\dagger = \hat v \hat k'^2 \hat v^\dagger,\qquad \hat D \hat D^\dagger = \hat w \hat k^2 \hat w^\dagger.
\end{equation}
Since $\hat k$ and $\hat k'$ are strictly positive and $\hat u$ and $\hat w$ are unitary matrices, we can define unitary
matrices $\hat x$ and $\hat y$ such that $\hat C$ and $\hat D$ have
the form
 \begin{equation}\label{C,D_dec_2}
 \hat C = \hat v \hat k'\hat x,\qquad \hat D = \hat w \hat k \hat y.
\end{equation}
 The remaining  unitary condition
$\hat A^{\dagger}\hat C+\hat B^{\dagger}{\hat D} = 0$ reads $\hat k\hat k'(\hat x +
\hat y) = 0$. Again, since $\hat k$ and $\hat k'$ are strictly positive this implies
$\hat y = - \hat x$, which fixes $\hat y$ in a
unique way.\newline
 The unitary matrix $\hat U$ then becomes:
\begin{equation}\label{U_1} \hat U = \begin{pmatrix}
  \hat v & 0 \\
  0 & \hat w
\end{pmatrix} \begin{pmatrix}
  \hat k & \hat k' \\
  \hat k' & -\hat k
\end{pmatrix} \begin{pmatrix}
  1 & 0 \\
  0 & \hat x
\end{pmatrix}.
\end{equation}

 In order to get a more symmetric form for $\hat U$, we introduce a unitary matrix
$\hat X$ such that $(\hat X^{\dagger})^{2} = \hat x$ and define the
matrices $\hat V = \hat v \hat X^{\dagger}$ and $\hat W = \hat w \hat X^{\dagger}$,
$\hat K = \hat X \hat k \hat X^{\dagger}$, $\hat K' = \hat X \hat k' \hat X^{\dagger}$. We
can then write the $4\times4$ unitary transformation operator
$\hat U$ as

\begin{equation}\label{Unitary_matrix}
\hat U =
\begin{pmatrix}
  \hat V & 0 \\
  0 & \hat W
\end{pmatrix}
\begin{pmatrix}
  \hat K & \hat K' \\
  \hat K' & -\hat K
\end{pmatrix}
\begin{pmatrix}
  \hat X & 0 \\
  0 & \hat X^{\dagger}
\end{pmatrix},
\end{equation}
in terms of the three unitary matrices $\hat V$, $\hat W$, $\hat X$ and the
non-negative hermitian matrices $\hat K$ and $\hat K' = \sqrt{1 -
\hat K^2}$. Since $\hat K$ and $\hat K'$ are strictly positive, this decomposition is unique, provided we fix the signs of the eigenvalues of
$\hat X = \sqrt{\hat x^{\dagger}}$ by some convention, for instance,
$(\hat X + \hat X^{\dagger})\geq 0$.

Having the unitary matrix given by (\ref{Unitary_matrix}) we can calculate the state of the overall system at later time as
\begin{equation}\label{final_state}
 \hat \Omega_f = \hat{U} \hat \Omega_0 \hat U^\dagger,
\end{equation}
where $\hat \Omega_0$ is given by (\ref{S+A_initial_state}).
The observables that can be simultaneously measured by means of the same apparatus on the state $\hat \Omega_f$ are the z-components of each spin. This corresponds to the following averages
\begin{eqnarray}\label{average}
\langle \hat s_z \rangle&=& \textnormal{tr} \left[ \hat \Omega_f \, \hat s_z \right],\nonumber\\
\langle \hat \sigma_z \rangle &=& \textnormal{tr} \left[ \hat \Omega_f \, \hat \sigma_z  \right],\nonumber\\
\langle \hat s_z \,\hat \sigma_z \rangle &=& \textnormal{tr} \left[ \hat \Omega_f \, (\hat s_z \,\hat \sigma_z)  \right].
\end{eqnarray}
We notice that the correlator $\langle \Omega_f \, (\hat s_z \,\hat \sigma_z) \rangle$ can be recoverd form the gathered data of $\hat \sigma_z$ and $\hat s_z$ via the number of coincidences.\newline
Inserting the  unitary time-evolution operator in the expectation values (\ref{average}) we get a linear relation between the gathered data of measurements of $\hat s_z$, $\hat \sigma_z$ and their correlator on one hand and the elements of the initial density matrix of $\textrm{S} $ given by $r_x$, $r_y$ and $r_z$ on the other hand. But before calculating the above mentioned expectation values, let us first parametrize the unitary time evolution operator $\hat U$.\newline

Since $\hat K$ is a Hermitian matrix  with $0 \leq \hat K \leq 1$, we can parametrize it as
\begin{equation}
 \label{K_parameter}
\hat K = \cos \theta \cos \phi + \sin \theta \sin \phi\, (\vec\chi \cdot \vec{\hat \sigma}),
\end{equation}
where $\vec \chi$  is a unit vector and $0 < \phi \leq \theta\leq \frac{\pi}{2} - \phi$. It is straightforward to see that $\hat K'$ is given by
\begin{equation}
 \label{K'_parameter}
\hat K' = \sin \theta \cos \phi - \cos \theta \sin \phi\, (\vec\chi \cdot \vec{\hat \sigma}),
\end{equation}

Since the initial overall density matrix $\hat \Omega_0$ is block diagonal, multiplication of the unitary matrix $\hat X$ by a phase factor does not affect $\hat U \hat \Omega_0 \hat U^\dagger$ although it modifies $\hat U$. Therefore, we can parametrize $\hat X$ as
\begin{equation}
 \label{X_parameter}
\hat X = e^{i \psi (\vec \xi \cdot \vec{\hat \sigma})} = \cos \psi + i (\vec \xi \cdot \vec{\hat \sigma}) \sin \psi,
\end{equation}
where $\vec\xi$ is a unit vector which we assume to be perpendicular to $\vec \chi$ for simplicity, and $0\leq \psi \leq \pi$.\newline
Parametrization of $\hat V$ and $\hat W$ can be done due to the fact that we are not interested in the off-diagonal block elements of $\hat \Omega_f$. In other words, the three expectation values (\ref{average}) do not require the determination of the off-diagonal elements of the overall density matrix and it would be sufficient to determine the action of $\hat V$ and $\hat W$ on the $\hat \sigma_z$:
\begin{equation}
 \label{V,W_parameter}
\hat V^\dagger \hat \sigma_z \hat V = \vec \eta \cdot \vec{\hat \sigma},\qquad \hat W^\dagger \hat \sigma_z \hat W = \vec \zeta \cdot \vec{\hat \sigma},
\end{equation}
where $\eta$ and $\zeta$ are three dimensional unit vectors.

Inserting the expression for $\hat \Omega_f$ from (\ref{final_state}) into (\ref{average}) using the parametrization introduced by (\ref{K_parameter})-(\ref{V,W_parameter}) yields
\begin{eqnarray}
&& \langle \hat s_z \rangle =\lambda \cos 2 \theta \cos 2 \phi + \lambda (\vec \chi \cdot \vec r) \sin 2 \theta \sin 2 \phi \cos 2 \psi\nonumber\\
&&+\left[(\vec \xi \times \vec \chi) \cdot \vec r\right] \sin 2 \theta \sin 2 \phi \sin 2 \psi ,\label{average_sz}
\end{eqnarray}
and 
\begin{eqnarray}
&&\langle \hat \sigma_z \rangle + \langle \hat s_z \,\hat \sigma_z \rangle = \lambda (\vec \chi \cdot \vec \eta) \sin 2 \theta \sin 2 \phi\nonumber\\
&&+ (\vec \chi \cdot  \vec \eta) (\vec \chi \cdot \vec r) (1 - \lambda \cos 2 \theta) (1 - \cos 2 \phi ) \cos 2 \psi\nonumber\\
&&+( \vec \chi \cdot  \vec \eta) \left[(\vec \xi \times \vec\chi )\cdot \vec r \right] ( \lambda - \cos 2 \theta) (1 - \cos 2 \phi) \sin 2 \psi\nonumber\\
&&+ (\vec \xi \cdot \vec \eta) (\vec \xi \cdot \vec r)(\cos 2 \phi + \lambda \cos 2 \theta) ( 1 - \cos 2 \psi)\nonumber\\
&&+(\vec \eta \cdot \vec r) (\cos 2 \phi + \lambda \cos 2 \theta) \cos 2 \psi \nonumber\\
&&+ \left[(\vec \xi \times \vec \eta) \cdot \vec r\right] (\lambda \cos 2 \phi + \cos 2 \theta) \sin 2 \psi.\label{average_sz+szsiz}
\end{eqnarray}
Finally, $\langle \hat \sigma_z \rangle - \langle \hat s_z \,\hat \sigma_z \rangle$ can be obtained by transforming $2 \theta$ to $2 \theta + \pi$ and replacing $\vec \eta$ with $\vec \zeta$ in (\ref{average_sz+szsiz}).

For the sake of simplicity, we assume that $\vec  \xi$ is the unit vector in the $x-$direction and that the unit vector $\vec \chi$ lies in the $y-$direction:
\begin{equation}
\vec \xi = (1, 0 , 0), \qquad \vec \chi = (0, 1 , 0),\qquad \vec \xi \times \vec \chi = (0,0,1).\label{xi,chi_max}
\end{equation}
Therefore the components of the two vectors $\vec \eta$ and $\vec \zeta$ on $\vec \xi$ and $\vec \chi$ can be defined as
\begin{eqnarray}
 &&\eta_x \stackrel{def}{=} \vec \xi \cdot \vec \eta,\qquad \eta_y \stackrel{def}{=}\vec \chi \cdot \vec \eta, \qquad \eta_z = [\vec \xi \times \vec \chi] \cdot \vec \eta\nonumber\\
&&\zeta_x \stackrel{def}{=} \vec \xi \cdot \vec \zeta,\qquad \zeta_y \stackrel{def}{=}\vec \chi \cdot \vec \zeta, \qquad \zeta_z = [\vec \xi \times \vec \chi] \cdot \vec \zeta.\nonumber\\\label{eta,zeta}
\end{eqnarray}

Within the above choice of the unit vectors we can relate the measured values of the population difference of the two energy-levels of $\textrm{A}$ and $\textrm{S}$ to the initial state of the system $\textrm{S}$ as
\begin{eqnarray}
 \begin{pmatrix}
  \langle \hat s_z \rangle\\
  \langle \hat \sigma_z \rangle\\
  \langle \hat s_z \,\hat \sigma_z \rangle
 \end{pmatrix} = {\cal C}
 \begin{pmatrix}
  r_x\\
  r_y\\
  r_z\\
 \end{pmatrix} + {\cal F},
\end{eqnarray}
where ${\cal C}$ is a $3\times 3$ matrix of coefficients the elements the elements of which are given by 
\begin{eqnarray}
c_{11} &=&0,\nonumber\\
c_{12} &=& \lambda \sin 2 \theta \sin 2 \phi \cos 2 \psi,\nonumber\\
c_{13} &=& \sin 2 \theta \sin 2 \phi \sin 2 \psi,\\\label{c}
c_{21}&=& (\eta_x + \zeta_x) \cos 2 \phi + \lambda (\eta_x - \zeta_x) \cos 2 \theta,\nonumber\\
c_{22} &=& (\eta_y + \zeta_y) \cos 2 \psi + \lambda (\eta_y - \zeta_y) \cos 2 \theta \cos 2 \phi \cos 2 \psi\nonumber\\
&-& \sin 2 \psi \left[ \lambda (\eta_z + \zeta_z) \cos 2 \phi + ( \eta_z - \zeta_z) \cos 2 \theta \right],\nonumber\\
c_{23}&=&\lambda (\eta_y + \zeta_y) \sin 2 \psi + (\eta_y - \zeta_y) \cos 2 \theta \cos 2 \phi \sin 2 \psi\nonumber\\
&+&\cos 2 \psi \left[ (\eta_z + \zeta_z) \cos 2 \phi  + \lambda (\eta_z - \zeta_z) \cos 2 \theta\right],\nonumber\\
c_{31}&=& (\eta_x - \zeta_x) \cos 2 \phi + \lambda (\eta_x + \zeta_x) \cos 2\theta,\nonumber\\
c_{32}&=& (\eta_y - \zeta_y) \cos 2 \psi + \lambda (\eta_y + \zeta_y) \cos 2 \theta \cos 2\phi \cos 2 \psi\nonumber\\
&-& \sin 2\psi\left[  \lambda (\eta_z - \zeta_z) \cos 2 \phi + (\eta_z + \zeta_z) \cos 2 \theta \right],\nonumber\\
c_{33}&=& \lambda (\eta_y - \zeta_y) \sin 2 \psi + (\eta_y + \zeta_y) \cos 2\theta \cos 2\phi \sin 2\psi\nonumber\\
&+& \cos 2 \psi \left[ (\eta_z - \zeta_z) \cos 2 \phi + \lambda(\eta_z + \zeta_z) \cos 2 \theta\right],\nonumber\\
\end{eqnarray}
 and ${\cal F}$ is the vector of constants given by:
\begin{eqnarray}
 {\cal F}&=&  \lambda\begin{pmatrix}
             \cos 2 \theta \cos 2 \phi\\
             (\eta_y - \zeta_y) \sin 2 \theta \sin 2 \phi\\ 
	      (\eta_y + \zeta_y) \sin 2 \theta \sin 2 \phi
            \end{pmatrix}.\label{f}
\end{eqnarray}

The elements of the initially unknown density matrix of $\textrm{S}$ which are encoded by $\vec r$ are related to these expectation values, so they can be recovered if and only if the determinant of the coefficient matrix ${\cal C}$ is non-zero. With some algebra we can calculate the determinant of the coefficient matrix, represented by $D$, as
\begin{eqnarray}
 \label{det}
\frac{8 \,D}{\sin 2 \theta \sin 2 \phi} &=&\frac{ (1 - \lambda^2)\sin 4 \psi  }{2}[ \left( \cos 2 \phi + \lambda \cos 2 \theta \right)\eta_x \zeta_y\nonumber\\
& - &\left( \cos 2 \phi - \lambda \cos 2 \theta \right) \eta_y \zeta_x ]\nonumber\\
&+&\eta_z\zeta_x(\cos 2 \phi - \lambda \cos 2 \theta) [ \lambda \cos 2\phi\nonumber\\
& +&\cos 2 \theta ( \lambda^2 \cos^2 2 \psi + \sin^2 2 \psi) ]\nonumber\\
&-& \eta_x \zeta_z (\cos 2 \phi + \lambda \cos 2 \theta) [ \lambda \cos 2\phi \nonumber\\
&- & \cos 2 \theta (\lambda^2 \cos^2 2 \psi + \sin^2 2 \psi ) ].
\end{eqnarray}

Thus the initial state of the system $\textrm{S}$ can be determined from $\langle \hat s_z \rangle$, $\langle \hat \sigma_z \rangle$ and $\langle \hat s_z \,\hat \sigma_z \rangle$ provided that the determinant $D$ is non-zero.\newline
In what follows we consider two limiting cases: {\it i}) when the assistant  is initially in a completely random state $(\lambda = 0)$, and {\it ii}) when it starts its evolution from a pure state, i.e. $\lambda = 1$. Then we maximize the value of $\Delta$ over the parameters of $\hat U$ and reconstruct the initial state of $\textrm{S}$. We also try to understand what kind of spin-spin interaction Hamiltonian will yield the maximum value of $D$ in each case.

\subsection{Assistant in completely disordered initial state}
Inserting $\lambda = 0$ in the expression for the determinant given by (\ref{det}) yields
\begin{eqnarray}
 \label{disorder_det}
D &=& \frac{1}{16} \sin 2 \theta \sin 4 \phi \sin 2 \psi [ \cos 2 \psi (\eta_x \zeta_y - \eta_y \zeta_x)\nonumber\\
 &+& \cos 2 \theta \sin 2 \psi (\eta_z \zeta_x + \eta_x \zeta_z) ].
\end{eqnarray}
It is clear that this determinant is maximized over the parameter $\phi$ if $\phi = \pm \frac{\pi}{8}$. Furthermore, the maximum of 
$D = \frac{1}{16} \sin 2 \theta \sin 2 \psi [ \cos 2 \psi (\eta_x \zeta_y - \eta_y \zeta_x)+ \cos 2 \theta \sin 2 \psi (\eta_z \zeta_x + \eta_x \zeta_z) ]$ over $\vec \eta$ and $\vec \zeta$ is reached when
\begin{eqnarray}
&&\vec \eta = \vec \zeta = (\frac{1}{\sqrt{2}}, 0,  \frac{1}{\sqrt{2}}).\label{eta,zeta_max}
\end{eqnarray}
Thus we have
\begin{equation}
 \label{max_delta_disordered_1}
D = \frac{1}{16} \sin 2 \theta \sin 2\psi \sqrt{1 - \sin^2 2\theta \sin^2 2\psi}.
\end{equation}
The determinant (\ref{max_delta_disordered_1}) reaches its maximum value $1/32$ for $\theta = \pi/ 8$ and $\psi = \pi /4$.  Such non-zero determinant guarantees the procedure of inversion and characterizing the initial state of the system. Inserting the above values of the parameters in the expressions for the expectation values of the $z$-component of spins, (\ref{average_sz}), (\ref{average_sz+szsiz}) we can reconstruct the initial density matrix of $\textrm{S}$:
\begin{eqnarray}\label{random_ini_S}
 r_x&=&2 \langle \hat \sigma_z \rangle,\nonumber\\
r_y&=&- 2 \langle \hat s_z \,\hat \sigma_z \rangle,\nonumber\\
 r_z&=&2 \langle \hat s_z \rangle.
\end{eqnarray}
We see that for a suitable choice of the evolution operator $\hat U$  it is possible to determine the initial state of a spin-$1/2$ system implying an assistant which is initially in completely disordered state.
\subsubsection{Construction of a Hamiltonian}
In order to build up a Hamiltonian such that the value of the determinant $\vert D \vert$ is maximized, we consider the following feasible criteria:
\begin{itemize}
 \item The interaction between $\textrm{A}$ and $\textnormal{S}$ is described by an anisotropic Heisenberg Hamiltonian
\begin{equation}
 \label{disordered_H1_def}
\hat{H}_{\textnormal{int}} = \sum_{i = x, y, z} J_i \hat \sigma_i \hat s_i,
\end{equation}
where $J_x, J_y$ and $J_z$ are the couplings for $x, y$ and $z$ components. This type of  Hamiltonian is employed to describe interacting spin systems in different areas such as spin chains, nuclear magnetic resonance (NMR), quantum optics, spintronics, etc.
 \item The possibly external fields acts symmetrically on $\textrm{A}$ and $\textrm{S}$.  Thus the total Hamiltonian can be written as
\begin{equation}
 \label{disordered_H2_def}
\hat H = \sum_{i = x, y, z} h_i (\hat \sigma_i + \hat s_i) + \hat H_{\textrm{int}},
\end{equation}
where $h_i$ are external fields in $x, y$ and $z$ direction.
\end{itemize}
Now let us try the following form:
\begin{equation}
 \label{dissordered_H_def}
\hat H = h_x (\hat \sigma_x + \hat s_x) + J_x (\hat \sigma_x s_x + \hat \sigma_z s_z) + J_y \hat \sigma_y s_y,
\end{equation}
which involves a partially anisotropic Heisenberg Hamiltonian and an external field acting on the $x-$direction.\newline
This Hamiltonian can be directly diagonalized. Its eigenvalues $E_k$ are
\begin{eqnarray}
 \label{disordered_H_eigenvalues}
&&E_1 = - 2 J_x - J_y,\qquad E_2 = J_y,\nonumber\\
&&E_3 = J_x - \beta,\nonumber\\
&& E_4 = J_x + \beta,
\end{eqnarray}
where we defined
\begin{equation}
 \beta = \sqrt{(J_x - J_y)^2 + 4 h_x^2}.
\end{equation}
The corresponding orthonormal eigenstates $\vert E_k \rangle$ , $k= 1, \ldots 4$, read
\begin{eqnarray}
 \label{disordered_H_eigenvectors}
&&\vert E_1 \rangle = \frac{1}{\sqrt{2}} (0, - 1, 1, 0 ),\nonumber\\
&&\vert E_2 \rangle = \frac{1}{\sqrt{2}} (- 1, 0 , 0 , 1),\nonumber\\
&&\vert E_3 \rangle = \frac{1}{\sqrt{2 (1 + \gamma^2)}} ( 1, \gamma, \gamma, 1),\nonumber\\
&&\vert E_4 \rangle = \frac{1}{\sqrt{2 (1 + \mu^2)}} ( 1, \mu, \mu, 1),
\end{eqnarray}
where $\gamma$ and $\mu$ are defined as
\begin{eqnarray}
 \label{gamma}
&&\gamma = \frac{J_y - E_4}{2 h_x},\\
&&\mu = \frac{J_y - E_3}{2 h_x}.\label{mu}
\end{eqnarray}
Now for $t = 1$ we can write
\begin{equation}
 \label{disordered_U}
\hat U = e^{- i \hat H} = \sum_{k = 1}^4 e^{- i E_k} \vert E_k \rangle \langle E_k \vert.
\end{equation}
The expectation value of any observable $\hat {\cal O}$ can be calculated as
\begin{eqnarray}
 &&\langle \hat {\cal O} \rangle = \textnormal{tr} \left[ \Omega_0 \hat U^\dagger \hat {\cal O} \hat U\right]= \sum_{k,l = 1}^4 \sum_{\alpha = x, y, z}\nonumber\\
&&\left[ e^{i (E_k - E_l)} \langle E_k \vert \hat {\cal O} \vert E_l \rangle \langle E_l \vert \hat \sigma_\alpha  \vert E_k \rangle r_\alpha\right],
\end{eqnarray}
where we have used (\ref{S+A_initial_state}). Calculation of $\langle \hat s_z \rangle$, $\langle \hat \sigma_z \rangle$ and theire correlator yields
\begin{eqnarray}
 \begin{pmatrix}
  \langle s_z \rangle\\
  \langle \hat \sigma_z \rangle\\
  \langle \hat s_z \,\hat \sigma_z \rangle
 \end{pmatrix} = 
{\cal P}\,
 \begin{pmatrix}
  r_x\\
 r_y\\
  r_z
 \end{pmatrix},
\end{eqnarray}
where ${\cal P}$ is a $3\times 3$ matrix whose determinant is 
\begin{eqnarray}\label{d}
 \textnormal{det}{\cal P} = \frac{(J_x - J_y)^2 h_x^2 \sin(4 J_x) \sin^4 \beta}{2 \beta^4}.
\end{eqnarray}
The determinant (\ref{d}) reaches its maximum value of $1/32$ for
\begin{eqnarray}
 \label{d_max}
J_x &=& \frac{\pi}{8},\nonumber\\
J_y&=& \frac{\pi ( 1 - \sqrt{8})}{8},\nonumber\\
h_x &=&\pm \frac{\pi}{2 \sqrt{8}}.
\end{eqnarray}
Thus the optimal measurements for the completely random initial state of the assistant are reachable with partially anisotropic Heisenberg Hamiltonian and one external magnetic field acting on both systems along the $x-$direction.\newline
We note that the physical reason of maximizing the determinant $D$ is to suppress the effect of the errors made during the measurement process.

%
 \subsection{Assistant in a pure initial state}
Considering the assistant starts its evolution from a pure state is equivalent to set $\lambda = 1$ in the general expression for the determinant given by (\ref{det}), which yields
\begin{equation}
 \label{det_pure}
D = \frac{1}{8} \sin 2\theta \sin 2 \phi (\cos^2 \phi - \cos^2 \theta ) (\eta_z \zeta_x - \eta_x \zeta_z).
\end{equation}
The maximum value of the determinant, $\vert D \vert = 1/12 \sqrt{3}$, in this case is reached when $\vec \eta$ and $\vec \zeta$ are perpendicular to each other
\begin{eqnarray}
 && \vec \eta  = (\frac{1}{\sqrt{2}}, 0, \frac{1}{\sqrt{2}}), \qquad \vec \zeta  = (\frac{1}{\sqrt{2}}, 0, -\frac{1}{\sqrt{2}}),
\end{eqnarray}
$\phi = \pm \frac{\pi}{4}$, and $\sin^2 (2 \theta) = 1/3$ while $\psi$ which determines a phase in the unitary operator remains an arbitrary parameter.\newline
Thus the initial state of $\textrm{S}$ can be determined as
\begin{eqnarray}\label{pure_ini_S_1}
 &&r_x = \sqrt{3} \langle \hat s_z \,\hat \sigma_z \rangle,\nonumber\\
&&r_y = \sqrt{3} \left(\cos 2 \psi \langle \hat s_z \rangle - \sin 2 \psi \langle \hat \sigma_z \rangle \right),\nonumber\\
&&r_z = \sqrt{3} \left(\sin 2 \psi \langle \hat s_z \rangle + \cos 2 \psi  \langle \hat \sigma_z \rangle \right).
\end{eqnarray}
If we choose the phase $\psi = \pi/4$, we get
\begin{eqnarray}\label{pure_ini_S_2}
 r_x &=& \sqrt{3} \langle \hat s_z \,\hat \sigma_z \rangle,\nonumber\\
 r_y &=& - \sqrt{3} \langle \hat \sigma_z \rangle,\nonumber\\
 r_z &=& \sqrt{3} \langle \hat s_z  \rangle.
\end{eqnarray}

Following the same line of arguments analogous to the random initial state of the assistant, a proper Hamiltonian corresponding to the optimal determinant of $1/12 \sqrt{3}$ is
\begin{eqnarray}
 \label{pure_H}
&&\hat H = \frac{1}{\sqrt{2}} \hat \sigma_x \hat s_x + \frac{1}{2} (\hat s_y \sin \alpha + \hat s_z),\nonumber\\
&& \sin \alpha = \sqrt{\frac{1}{2} \left( 1 - \frac{1}{\sqrt{3}}\right)}.
\end{eqnarray}
This Hamiltonian represents an Ising interaction between $\textrm{A}$ and $\textrm{S}$ together with an action of a magnetic field on the assistant.

\section{\label{light_assistant}Light as an Assistant}
In this section we introduce another type of assistant. Specifically we show that the unknown density matrix of an ensemble of two-level systems (atom or spin) can be determined via interaction with a single mode of the electromagnetic field. In this case the type of the interaction Hamiltonian is fixed from the beginning. The atom-field interaction is studied within the \index{Jaynes-Cummings model(JCM)}Jaynes-Cummings model (JCM)~\cite{JCM},\cite{Scully1997}. Thus the unitary time-evolution operator is known. We choose two commuting observables of the overall system and show that the initial state of $\textrm{S}$ is linearly dependent on the expectation values of the two observables and their correlator.
In this case, the unknown state of the spin can be characterized by repeated measurement of two commuting observables: the population difference of the atoms $\hat{\sigma}_z$, and the photon number of the field $\hat{a}^\dagger \hat{a}$. This measurement supplies three averages: $\langle \hat{\sigma}_z \rangle$, $\langle  \hat{a}^\dagger \hat{a} \rangle$, and $\langle  \hat{\sigma}_z \,\hat{a}^\dagger \hat{a} \rangle$, which will be linearly related to the elements of the initial density matrix of the ensemble of the two-level atoms. (Note that
since $\hat{\sigma}_z$ and $\hat{a}^\dagger \hat{a}$ commute,
$\langle\hat{\sigma}_z\,\hat{a}^\dagger \hat{a}\rangle$ is recovered from the
measurement data of $\hat{\sigma}_z$ and $\hat{a}^\dagger \hat{a}$ via the number of
coincidences. This is similar to previous section, where $\langle \hat s_z \,\hat \sigma_z \rangle$ could be determined from measurements of $\hat s_z$ and $\hat \sigma_z$.)
\subsection{Interaction Hamiltonian}

The Jaynes-Cummings Hamiltonian for the interaction of a two-level system with a single mode field  reads~\cite{Mehmani2008}
\begin{equation}
\label{JC_Hamiltonian}
\hat{H} = \hbar \omega
\hat{\sigma}_z + \hbar \nu \hat{a}^{\dagger} \hat{a} + \hbar g
(\hat{\sigma}_+ \hat{a} + \hat{\sigma}_- \hat{a}^{\dagger}),
\end{equation}
where we restricted ourselves in the dipole approximation~\cite{Scully1997}. $\omega$ and $\nu$ are the atom and field frequencies, respectively. $\hat a$ and $\hat a^\dagger$ are field annihilation and creation operators, while $\hat{\sigma}_+$ and $ \hat{\sigma}_-$ are the raising and lowering spin operators, and $g$ is the coupling constant.
We shall denote
\begin{equation*}
\Delta\stackrel{def}{=} \omega - \nu,
\end{equation*}
for the \index{Detuning parameter}detuning parameter. For our future purposes we
note that $\Delta $ is a tunable parameter.

The time evolution operator of the Jaynes-Cummings (JC) Hamiltonian, $\hat U(t)$,  can be calculated in an exact manner [see~ \cite{Mehmani2008} and references therein.].

In the eigenbasis of the two-level system $\hat{U}(t)$ reads:
\begin{eqnarray}\label{ch2,eq:U_fin}
&&\hat{U}(t) = e^{- i \nu t (\hat{a}^{\dagger} \hat{a} + \frac{1}{2})}\nonumber\\
&&\times \left( \cos [t \sqrt{\hat{\varphi} + g^2}] - i \frac{\Delta}{2}\, \frac{\sin [t  \sqrt{\hat{\varphi} + g^2}]}{\sqrt{\hat{\varphi} + g^2}}\right) |+\rangle\langle+|\nonumber\\
&&- i g e^{- i \nu t (\hat{a}^{\dagger} \hat{a} + \frac{1}{2})} \frac{\sin [t  \sqrt{\hat{\varphi} + g^2}]}{\sqrt{\hat{\varphi} + g^2}} \,\hat{a} |+\rangle\langle-|\nonumber\\
&& -i g e^{- i \nu t (\hat{a}^{\dagger} \hat{a} - \frac{1}{2})} \frac{\sin t \sqrt{\hat{\varphi}}} {\sqrt{\hat{\varphi}}} \,\hat{a}^{\dagger} |-\rangle\langle+|\nonumber\\
&&+ e^{- i \nu t (\hat{a}^{\dagger} \hat{a} - \frac{1}{2})} \left( \cos t \sqrt{\hat{\varphi}} + i \frac{\Delta}{2}\, \frac{\sin t \sqrt{\hat{\varphi}}}{\sqrt{\hat{\varphi} }}\right) |-\rangle\langle-|,\nonumber\\
\end{eqnarray}
where $\vert \pm \rangle$ are the eigenstates of $\hat{\sigma}_z$ with eigenenergies $E_{\pm}$.

The unitarity of $\hat{U}(t)$ is satisfied because of the identities

\begin{eqnarray}\label{ch2,eq:unitary_condition}
&&\frac{\sin \left[t\,\sqrt{\hat{\varphi} + g^2}\right]}{\sqrt{\hat{\varphi} +
g^2}}\; \hat{a} = \hat{a}\; \frac{\sin\left[
t\,\sqrt{\hat{\varphi} } \right] }{\sqrt{\hat{\varphi}}} ,\nonumber\\
&&\cos \left[t\, \sqrt{\hat{\varphi} + g^2}\right]\; \hat{a} = \hat{a}\;
\cos \left[t \sqrt{\hat{\varphi}}\right].
\end{eqnarray}

Having $\hat{U}(t)$ at hand, we can calculate any property of $S+A$, the overall system, we wish.
We consider the most general form of the initial state for the atom. This is described by some general mixed density matrix
$ \hat{\rho}$ given by (\ref{S_initial_state}).

For the assistant, we shall assume that the single cavity mode starts its
evolution from a coherent state with a known parameter $\alpha$:
\begin{equation}\label{ch2,eq:field_state}
|\alpha\rangle = e^{- |\alpha|^2/2} \sum_{n=0}^{\infty} \frac{\alpha^n}{\sqrt{n!}}\, |n\rangle,
\end{equation}
where $|\alpha\rangle$ is the eigenvector of the annihilation operator $\hat{a}$,
\begin{eqnarray*}
\hat{a}|\alpha\rangle = \alpha|\alpha\rangle,
\end{eqnarray*}
and where $|n\rangle$ is the eigenvector of the photon number operator $\hat{a}^{\dagger} \hat{a}$,
\begin{equation*}
\hat{a}^{\dagger} \hat{a}|n\rangle = n|n\rangle.
\end{equation*}
The assumption (\ref{ch2,eq:field_state}) on the initial state of the field is natural since these are
the kinds of fields produced by classical currents \cite{Glauber1965}, and also, to a
good approximation, by sufficiently intense laser fields.\newline
We assume the system and the assistant are initially separated and do not interact with each other. As a result, the overall initial density matrix is factorized,
\begin{equation}\label{ch2,eq:rho(0)}
\hat{{\cal D}}(0) = \hat{\rho}\,|\alpha\rangle \langle \alpha |,
\end{equation}
where the initial state of the system in the eigen-basis of $\hat{\sigma}_z$ reads
\begin{eqnarray}\label{ch2,eq:S_state_+-_representation}
&&\hat{\rho} = \left(\frac{1}{2} + r_z\right) \vert + \rangle \langle + \vert +\left( r_x - i r_y \right) \vert + \rangle \langle - \vert\nonumber\\
&&+\left( r_x + i r_y \right) \vert - \rangle \langle + \vert
+ \left(\frac{1}{2} - r_z \right) \vert - \rangle \langle - \vert.
\end{eqnarray}
The initial state of the field is given by
\begin{equation}\label{ch2,eq:field_state_n_representation}
 \vert \alpha \rangle \langle \alpha \vert = e^{- \vert \alpha \vert^2} \,\sum_{n = 0 }^\infty \sum_{m = 0 }^{\infty} \frac{\alpha^n {\alpha^*}^m}{\sqrt{n!} \sqrt{m!}}\vert n \rangle \langle m \vert.
\end{equation}
\begin{figure*} [t!]
      \centering
      \subfigure[$\,\,\bar n = 2, \Delta=10 kHz.$]
      {\label{fig:a}\includegraphics[width=.4\textwidth]{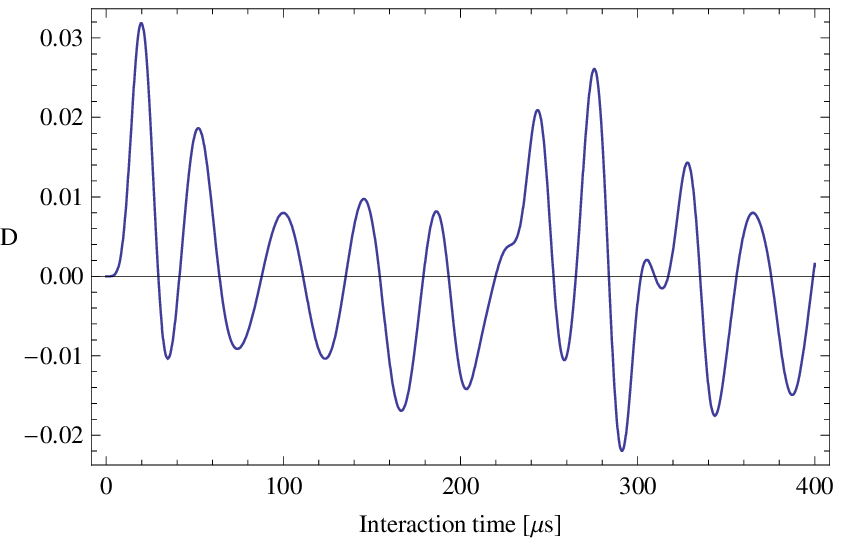}}
    \hspace{0.3cm}
      \subfigure[$\,\,\bar n =2, \Delta=100 kHz.$]
      {\label{fig:b}\includegraphics[width=.4\textwidth]{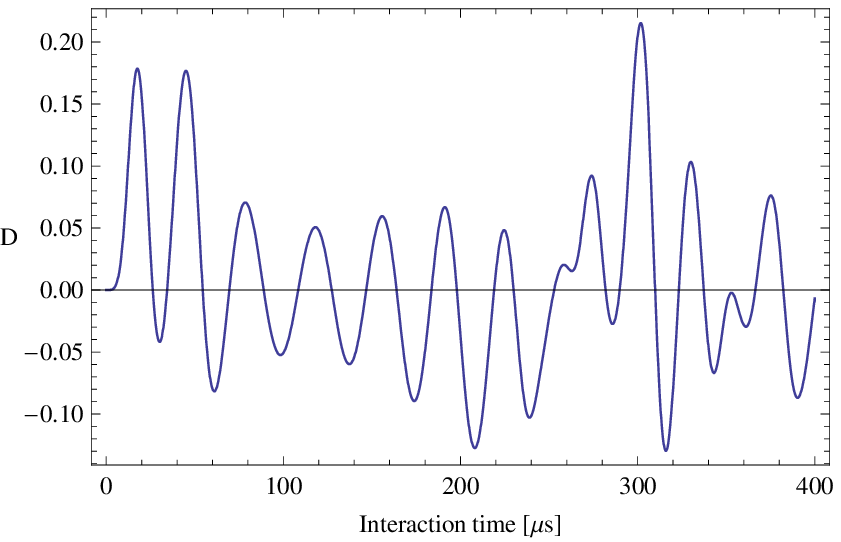}}
     \hspace{0.3cm}
      \subfigure[$\,\,\bar n = 5, \Delta = 10KHz.$]
     {\label{fig:c}\includegraphics[width=.4\textwidth]{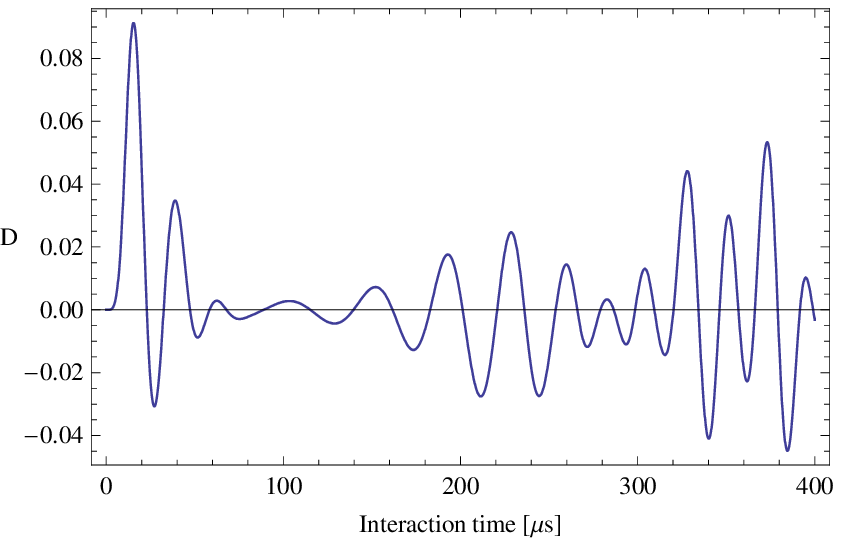}}
  \centering
     \subfigure[$\,\,\bar n = 5, \Delta=100 kHz$]
      {\label{fig:d}\includegraphics[width=.4\textwidth]{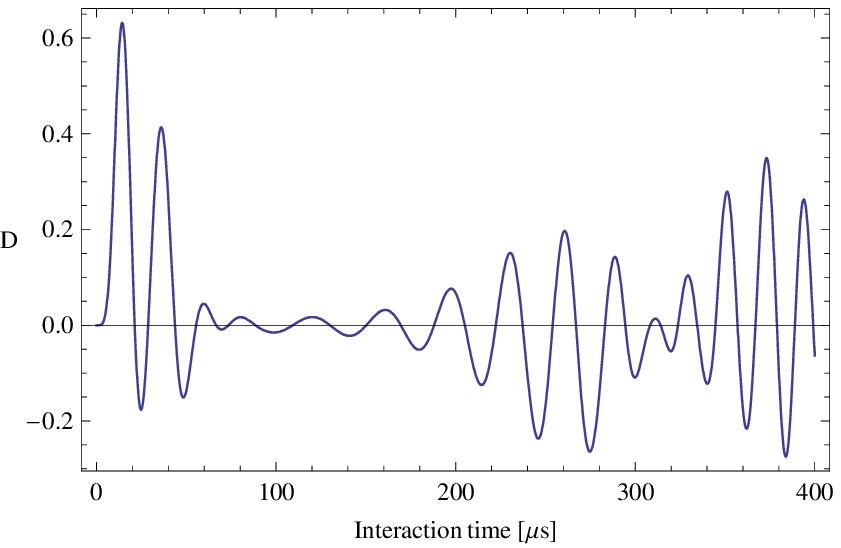}}
     \hspace{0.3cm}
  \centering
      \subfigure[$\,\,\bar n = 10, \Delta=10 kHz$]
      {\label{fig:e}\includegraphics[width=.4\textwidth]{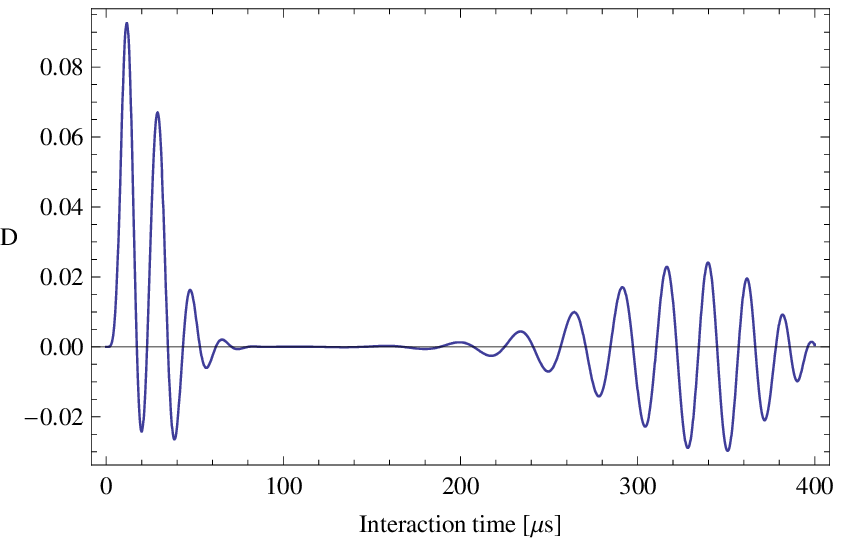}}
     \hspace{0.3cm}
  \centering
      \subfigure[$\,\,\bar n = 10, \Delta=100 kHz$]
      {\label{fig:f}\includegraphics[width=.4\textwidth]{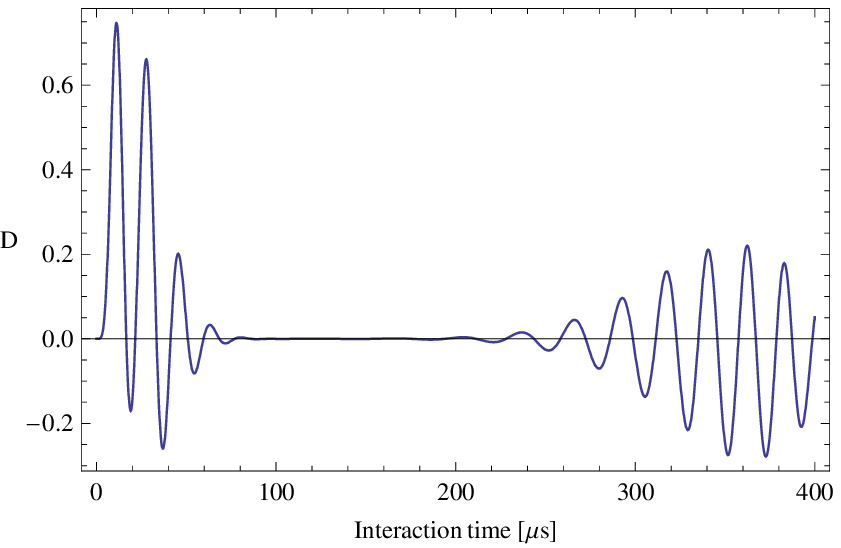}}
     \hspace{0.3cm}
 \caption{The dynamics of the determinant of the matrix ${\cal M}$  in the Jaynes-Cummings model for different values of\\ the mean photon number in the cavity, $\bar n = 2, 5, 10$ with two detuning parameters: $\Delta = 10KHz$ and $\Delta = 100 KHz$. The figures are cited from reference~\cite{Mehmani2008}. }
\label{fig:det}
 \end{figure*}
The state of the system + assistant at later time $t$ can be calculated with the help of the unitary
operator (\ref{ch2,eq:U_fin})
\begin{equation}
\label{ch2,eq:final_overall_state}
\hat{{\cal D}}(t) = \hat{U}(t)\, \hat{\rho}(0)\, \hat{U}^{\dagger}(t).
\end{equation}
Then the expectation value of any observable  $\hat{\cal{O}}$ of the overall system at time $t$
is
\begin{equation}\label{ch2,eq:expectation}
\langle\hat{\cal{O}}\rangle = \textnormal{tr}\left[ \hat{{\cal D}}(t) \hat{\cal{O}} \right].
\end{equation}
The next step is to calculate the two commuting observables with which we can build up the initial state of the atom. Using (\ref{ch2,eq:U_fin}), (\ref{ch2,eq:S_state_+-_representation}), (\ref{ch2,eq:field_state_n_representation}), and (\ref{ch2,eq:final_overall_state}), the atom population difference $\langle \hat{\sigma}_z \rangle_t$ reads
\begin{eqnarray}\label{ch2,eq:atom_population_difference}
&&\langle\hat{\sigma}_z\rangle_t = \frac{g^2}{2} \sum_{n=0}^{\infty} (n+1)(c_{n+1} - c_n) \frac{\sin^2\left(\Omega_n t/2\right)}{(\Omega_n t/2)^2}\nonumber\\
&& +4 \, g \,r_x \,\sum_{n=0}^{\infty}  c_n \frac{ \sin \left( \Omega_{n} t /2\right)} {\Omega_{n}} \Im\{ \chi_n(t)\}\nonumber\\
&&+ 4 \,g\, r_y \,\sum_{n=0}^{\infty}  c_n \frac{\sin \left( \Omega_{n}t /2\right)} {\Omega_{n}}\Re\{\chi_n(t)\}\nonumber\\
&&+ r_z\, \left\{1 -g^2\sum_{n=0}^{\infty}(n+ 1) (c_{n+1} + c_n )\frac{ \sin^2 \left( \Omega_{n} t /2\right)} {(\Omega_{n}/2)^2}\right\},
\nonumber\\
\end{eqnarray}
where $r_x, r_y$, and $r_z$
are the unknown elements of the initial atom density matrix, which we want to find out, $\Re$ and $\Im$ stand for the real and the imaginary parts of the argument in front of them, respectively. The parameters $\chi_n(t)$, $c_n$ are defined as
\begin{equation}\label{ch2,eq:chi_def}
\chi_n(t) \stackrel{def}{=} \alpha\,\left[\cos \left( \frac{\Omega_{n}\,t }{2}\right) + i \Delta\,\frac{\sin \left( \Omega_{n} t /2\right)} {\Omega_{n}}\right],
\end{equation}
and
\begin{equation}\label{ch2,eq:cn_def}
c_n \stackrel{def}{=} e^{- |\alpha|^2}\,\frac{\alpha^{2n}}{n!},
\end{equation}
where the corresponding Rabi frequency, $\Omega_n$, is defined as
\begin{equation}\label{ch2,eq:Omega_def}
{\Omega_n}\stackrel{def}{=}\sqrt{ 4 (n + 1) g^2 + \Delta^2}.
\end{equation}
The average number of photons in the cavity, $\langle \hat{a}^\dagger \hat{a} \rangle_t$, can be calculated in a similar way
\begin{eqnarray}\label{ch2,eq:photon_number}
&&\langle \hat{a}^\dagger \hat{a} \rangle_t = \sum_{n=0}^{\infty}n c_n \nonumber\\
&&- \frac{g^2}{2} \sum_{n=0}^\infty (n+1) (c_{n+1} - c_n) \frac{\sin^2(\Omega_n t/2)}{(\Omega_n/2)^2}\nonumber\\
&&- 4 \, g \, r_x\, \sum_{n=0}^{\infty}  c_n \frac{ \sin \left( \Omega_{n}t/2\right)} {\Omega_{n}} \Im\{ \chi_n(t)\}\nonumber\\
&&- 4\,  g\,  r_y\, \sum_{n=0}^{\infty}  c_n \frac{\sin \left( \Omega_{n}t /2\right)} {\Omega_{n}}\Re\{\chi_n(t)\}\nonumber\\
&&+ g^2 \, r_z\, \sum_{n=0}^\infty (n+1) (c_{n+1} + c_n) \frac{\sin^2(\Omega_n t/2)}{(\Omega_n /2)^2}.
\end{eqnarray}
The correlator of the two observables, $\langle \hat{\sigma}_z \,\hat{a}^\dagger \hat{a} \rangle_t$, which amounts to the number of coincidences, reads
\begin{eqnarray}\label{ch2,eq:correlator}
&&\langle \hat{\sigma}_z \,\hat{a}^\dagger \hat{a} \rangle_t =\nonumber\\
&& \frac{g^2}{4}\sum_{n=0}^\infty (n+1) \left[ (2 n +3) c_{n+1} - (2 n + 1) c_n\right]\times \nonumber\\
&& \times\frac{\sin^2(\Omega_n t/2)}{(\Omega_n /2)^2}\nonumber\\
&& + 2\, g\, r_x\, \sum_{n=0}^{\infty}  c_n  ( 2 n + 1 ) \frac{ \sin \left( \Omega_{n} t /2\right)} {\Omega_{n}} \Im\{ \chi_n(t)\}\nonumber\\
&& + 2\, g\, r_y\,\sum_{n=0}^{\infty}  c_n (2 n + 1 ) \frac{\sin \left( \Omega_{n} t /2\right)} {\Omega_{n}}\Re\{\chi_n(t)\}\nonumber\\
& + & r_z\, \sum_{n=0}^{\infty}\{n c_n  - \frac{(n+1) g^2}{2}\times\nonumber\\
&&\times \left[ (2 n + 3) c_{n+1} + (2 n + 1)  c_n \right] \frac{\sin^2(\Omega_n t/2)}{(\Omega_n /2)^2} \}.\nonumber\\
\end{eqnarray}
Expectedly, these three quantities, i.e.,  the atom population difference $\langle \hat{\sigma}_z \rangle_t$, the average number of photons $\langle \hat{a}^\dagger \hat{a} \rangle_t$, and the correlator of these two observables $\langle \hat{\sigma}_z \,\hat{a}^\dagger \hat{a} \rangle_t$ are linearly related to the three unknown parameters $r_x $, $r_y$, $r_z$ of the initial atom density matrix:
\begin{equation}\label{ch2,eq:matrix_equation}
\begin{pmatrix}
  \langle \hat{\sigma}_z \rangle_t\\
  \langle \hat{a}^\dagger \hat{a}\rangle_t  \\
  \langle\hat{\sigma}_z \,\hat{a}^\dagger \hat{a}\rangle_t
\end{pmatrix} = \cal{M} \begin{pmatrix}
  r_x  \\
 r_y  \\
 r_z
\end{pmatrix} + {\cal B}, \qquad
{\cal B} =
 \begin{pmatrix}
 b_1\\
 b_2\\
 b_3
 \end{pmatrix}.
\end{equation}
The elements of the $3 \times 3$ matrix ${\cal M}$ and the vector ${\cal B}$ are read off from Eqs. (\ref{ch2,eq:atom_population_difference}) -- (\ref{ch2,eq:correlator}). They depend
on the parameter $\alpha$ of the initial assistant state, on the detuning parameters $\Delta$, coupling $g$ of the
JC Hamiltonian, and on the interaction time $t$. Thus, if the matrix ${\cal M}$ is non-singular, i.e., the determinant of $\cal{M}$ is not zero,
one can invert $\cal{M}$ and express the unknown parameters of the initial atom density matrix via
known quantities. Although the elements of $\cal{M}$ are complicated, the
determinant itself is much simpler. It takes the explicit form
\begin{eqnarray}
\label{ch2,eq:determinant}
&&D(t)\stackrel{def}{=}{\rm det}[{\cal M}] = 4 \Delta\, g^2 e^{- 2 |\alpha|^2} \nonumber\\
&&\times \sum_{n=0}^{\infty}\sum_{m=0}^{\infty} \frac{|\alpha|^{2(n+m+1)}}{n! m!} (n-m) \times\nonumber\\
&&\left[\frac{\sin^2\left( \Omega_{n} t/2\right) \,\,\sin  \Omega_{m}\,t}{\Omega_{n}^2 \Omega_{m}} - \frac{\sin^2 \left(\Omega_{m}t/2\right)\,\,\sin \Omega_{n}\,t}{\Omega_{m}^2 \Omega_{n} }\right].\nonumber\\
\end{eqnarray}
At the
initial time $t=0$, the determinant
$D(0)$ is naturally zero, since the initial state of the
overall system is factorized. Note that $D(t)=0$ for
$\Delta=0$.  Aquino {\it et al.}~\cite{Aquino2007}  showed in the resonance case this scheme can be implied for the spin-$\frac{1}{2}$ systems but not for a quasi-spin-$\frac{1}{2}$ system. Thus some non-zero detuning is crucial for the present
scheme of the state determination of any two-level system.  The value of $D(t)$ changes by varying
the detuning parameter $\Delta$. It is seen in Figs.~\ref{fig:a}--\ref{fig:f} that for a non-zero
detuning, $\Delta\neq 0$, the determinant $(t)$ is non-zero for a certain
period $t>0$. (Obviously $D(t)$ is zero when there is no photon in the cavity.)
On the other hand, large $D(t)$ suggests that the state of the atom and the field are entangled~\cite{Aquino2008}.\newline
\begin{figure*}[t!]
  \centering
      \subfigure[$\,\,\Delta=10 kHz, \sigma= 0.1 \mu s$]
      {\label{fig:ave1}\includegraphics[width=.4\textwidth]{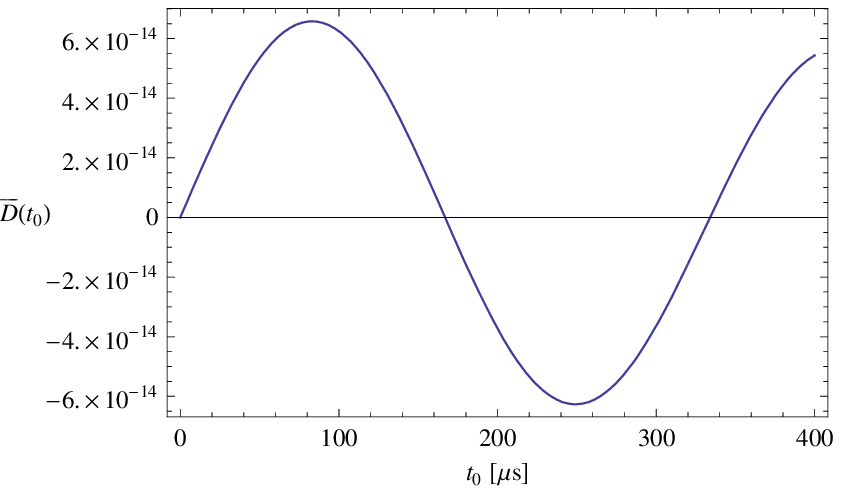}}
     \hspace{0.3cm}
  \centering
      \subfigure[$\,\,\Delta=100 kHz, \sigma= 0.01 \mu s$]
      {\label{fig:ave2}\includegraphics[width=.4\textwidth]{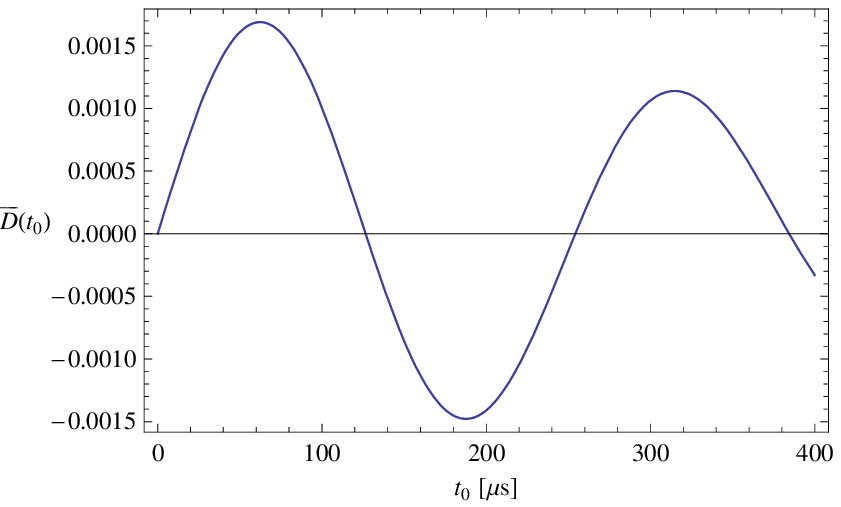}}
     \hspace{0.3cm}
 \caption{The time averaged determinant $\bar{D}$ in the Jaynes-Cummings model
 as a function of $t_0$ when the mean photon number in the cavity is $\bar{n} = 2$, and $g = 50kHz$ for different values of $\Delta$ and $\sigma$; see Eqs.~(\ref{ch2,eq:probability})-(\ref{ch2,eq:averaged_determinant_w}). The figures are cited from reference~\cite{Mehmani2008}. }
\label{fig:ave}
\end{figure*}
Comparing figures Fig.~\ref{fig:a} and Fig.~\ref{fig:c}
we see that although higher initial photon numbers $\bar{n}$ lead to
bigger values for the determinant, they cause rapid oscillations in the
value of the determinant. This makes the measurement process more difficult. (Note
in this context that the determinant depends on the absolute value of
$\alpha$ and $\bar{n} = |\alpha|^2$ is the average number of photons.)\newline
If the average number of photons $\bar{n}=|\alpha|^2$ in the initial
state of the field is sufficiently large, the determinant is nearly zero for
intermediate times; see Figs.~\ref{fig:e} and \ref{fig:f}. 
The reason for this collapse is apparent from
(\ref{ch2,eq:determinant}) and has the same origin as the collapse of the
atomic population difference well known for the JCM~\cite{Shore1993}. Each
term in the right hand side of (\ref{ch2,eq:determinant}) oscillates with a different
frequency. With time these oscillations get out of phase and $D(t)$
vanishes (collapses).  However, since the number of relevant
oscillations in $D(t)$ is finite, they partially get in phase for later
times producing the revival of $D(t)$, as seen in the
Figs.~\ref{fig:e} and \ref{fig:f}.\newline
It is seen that $D(t)$ does not depend on separate frequencies $\omega$ and
$\nu$ of the two-level system and the field, only their difference
$\Delta=\omega-\nu$ is relevant. This is due to the choice of the
measurement basis|see the left hand side of (\ref{ch2,eq:matrix_equation})|that involves quantities
which are constants of motion for $g\to 0$. Comparing the figures
Fig.~\ref{fig:a} with Fig.~\ref{fig:b}, Fig.~\ref{fig:c} with
Fig.~\ref{fig:d}, and Fig.~\ref{fig:e} with Fig.~\ref{fig:f}
one observes that the value of the highest peak of $D$ increases by an
order of magnitude when the detuning parameter changes from 10kHz to
100kHz.  Note that in Eq.~(\ref{ch2,eq:determinant}) for the determinant $D(t)$ the
contribution from the diagonal $n=m$ matrix elements of the assistant
initial state $|\alpha\rangle\langle \alpha|$ cancels out. Thus, it is
important to have an initial state of the assistant with non-zero
diagonal elements in the $\{ |n\rangle\}$ basis.\newline
Since the determinant $D(t)$ is not
zero for a realistic range of the parameters, the
initial unknown state of the two level system can be determined by
specifying the average atom population difference
$\langle\hat{\sigma}_z\rangle_t$, the average number of photons $\langle\hat{a}^\dagger
\hat{a}\rangle_t$, and their correlator $\langle\hat{\sigma}_z \,\hat{a}^\dagger
\hat{a}\rangle_t$.
\subsection{\label{ch2,sec:initial_state,subsec:random_t}Random interaction time}
We saw in the previous sections that the success of the presented scheme is to a large extent
determined by the ability to select properly the interaction time $t$,
since this ultimately should ensure a non-zero (and sufficiently
large) determinant $D(t)$ since a small determinant will amplify
numerical errors.\newline
We notice that the expectation value of an observable ${\cal O}$, as it is described in section I, is the ensemble average of an observable. This is mathematically denoted by $\textnormal{tr} \left[\hat \rho \hat {\cal O}\right]$. Now we have to take into account that the repeated measurement of counting the number of the photon in the cavity and the population difference of atoms are performed at a random $t$ in each set of measurement, which obeys the Gaussian distribution. Thus we have to perform a time-average in the relevant time window as well. Since we just want to get a rough estimation about the consequence of such way of measurement on the value of the determinant, we avoid the tedious time-averaging calculation of $\langle \hat a^\dagger \rangle$, $\langle \hat \sigma_z \rangle$, and their correlation by making a shortcut and perform the time-averaging of the determinant itself.\newline
To quantify the robustness of the presented scheme
it is reasonable to assume that
there is no perfect control in choosing the interaction time.
To this end let us assume
that the interaction time $t$ is a random, Gaussian distributed
quantity centered at $t_0$ with a
dispersion $\sigma$ and that an ensemble of measurements is performed to map out this spread. The corresponding probability distribution $P(t)$ of
thus reads
\begin{equation}\label{ch2,eq:probability}
P(t)=\frac{1}{2\pi\sigma}\, e^{-(t-t_0)^2/(2\sigma)}.
\end{equation}
Since n order to get a roughly estimation about the influence of the presence of such randomness during the measurement procedure on the value of the determinant we make a short cut and time
average the determinant $D(t)$ over the same probability distribution
\begin{eqnarray}
\label{ch2,eq:averaged_determinant}
&&\overline{D}(t_0)
 =4 \Delta\, g^2 e^{- 2 |\alpha|^2}
\sum_{n=0}^{\infty}\sum_{m=0}^{\infty} \frac{|\alpha|^{2(n+m+1)}}{n!
m!} \times\nonumber\\
&&\times(n-m)
\left[
w(\Omega_n,\Omega_m;t_0)-w(\Omega_m,\Omega_n;t_0)
\right],\nonumber\\
\end{eqnarray}
where
\begin{eqnarray}
\label{ch2,eq:averaged_determinant_w}
w(\Omega_n,\Omega_m;t_0)&=&
\frac{1}{4\Omega_n^2\Omega_m}\,\{2 e^{-\frac{\sigma}{2}{\Omega_m}^2 }\, \sin[t_0 \Omega_m ] \nonumber\\ &-&e^{-\frac{\sigma}{2}(\Omega_m+\Omega_n)^2} \sin[t_0 (\Omega_m+\Omega_n)] \nonumber\\
&-&e^{-\frac{\sigma}{2}(\Omega_m-\Omega_n)^2} \sin[(\Omega_m-\Omega_n) t_0]\}.\nonumber\\
\end{eqnarray}
It is seen that the oscillations of $D(t)$ turn after averaging into
exponential factors $e^{-\frac{\sigma^2}{2}(\Omega_m \pm \Omega_n)^2}$
and $e^{-\frac{\sigma^2}{2}\Omega_m^2}$ in (\ref{ch2,eq:averaged_determinant}, \ref{ch2,eq:averaged_determinant_w}),
due to which the averaged determinant $\overline{D}(t_0)$ gets
suppressed for a sufficiently large ``indeterminacy'' $\sigma$. This suppression is
illustrated in Fig.~\ref{fig:ave1} and Fig.~\ref{fig:ave2}.
By comparing Fig.~\ref{fig:ave1} and Fig.~\ref{fig:ave2} we realize that when the dispersion $\sigma$ grows by one order of magnitude, the value of the averaged determinant drops dramatically.

\section{\label{con}Conclusion}
It is important to implement the single-apparatus tomography for a
situation with a physically transparent measurement base and with a
realistic system-assistant interaction. Here we carried out this program for a two-level atom (system).
The non-commutating elements of the density matrix of a quantum system can be determined by simultaneous measurements of commuting observables. This can be done by introducing another system which its initial state is known, called assistant.  The two systems interact and then by performing repeated measurements of one observable belonging to the system of interest and the other one to the assistant, one is able to count the events in repeated experiments. This yields to a one-to-one correspondence between the initial density matrix and the collected data. 
We displayed several examples. In the first series of examples we considered the assistant to be a two-level system with a known initial state. We showed that the full initial density matrix of the system can be determined via simultaneous measurements of the occupation probabilities for the energy-levels of two systems. We discussed two different initial states for the assistant and showed that this procedure is feasible even when the assistant starts its evolution from completely disordered state. \newline
The dynamical processes which properly yield the determination of the initial state of the system can be described by an Ising or Heisenberg interaction Hamiltonian, regarding to the situation.\newline
We also showed that another way of reconstructing the state of a two-level system is to let it interact with a specific type of environment, namely, a coherent single mode of an electromagnetic field. In this case the interaction Hamiltonian is described by the Jaynes-Cummings Hamiltonian. In this sitation, the measurement of the simplest set of observables related to the energies of
the atom and field yield the determination of the unknown initial state of the atom.
\bibliography{refs}

\begin{thebibliography}{10}

\bibitem{BALLENTINE1970}
L.~E. Ballentine.
\newblock The statistical interpretation of quantum mechanics.
\newblock {\em Rev. Mod. Phys.}, 42(4):358--381, Oct 1970.

\bibitem{Muynck2002}
W.~de~Muynck.
\newblock {\em Foundations of Quantum Mechanics, an Empricist Approach}.
\newblock Kluwer Academia, Dordrecht, 2002.

\bibitem{Balian1989}
Roger Balian.
\newblock On the principles of quantum mechanics and the reduction of the wave
  packet.
\newblock {\em American Journal of Physics}, 57(11):1019--1027, 1989.

\bibitem{Nielsen2000}
M.~A. Nielsen and I.~L. Chuang.
\newblock {\em Quantum Computation and Quantum Information}.
\newblock Cambridge University Press, Cambridge, England, 2000.

\bibitem{Pasquinucci2000}
H.~Bechmann-Pasquinucci and W.~Tittel.
\newblock Quantum cryptography using larger alphabets.
\newblock {\em Phys. Rev. A}, 61(6):062308, May 2000.

\bibitem{Allahverdyan2004}
Th. M.~Nieuwenhuizen A.~E.~Allahverdyan, R.~Balian.
\newblock Determining a quantum state by means of a single apparatus.
\newblock {\em Phys. Rev. Lett.}, 92:120402, 2004.

\bibitem{D'Ariano2002}
G.~M. D'Ariano.
\newblock Universal quantum observables.
\newblock {\em Physics Letters A}, 300(1):1 -- 6, 2002.

\bibitem{Mehmani2008}
B.~Mehmani, A.~E. Allahverdyan, and Th.~M. Nieuwenhuizen.
\newblock Quantum-state tomography using a single apparatus.
\newblock {\em Phys. Rev. A}, 77(3):032122, Mar 2008.

\bibitem{Peng2007}
J.~Du X.~Peng and D.~Suter.
\newblock Measuring complete quantum states with a single observable.
\newblock {\em Physical Review A (Atomic, Molecular, and Optical Physics)},
  76(4):042117, 2007.

\bibitem{Aquino2007}
G.~Aquino and B.~Mehmani.
\newblock Simultaneous measurement of non-commuting observables.
\newblock In B.~Mehmani M. J.~Aghdami Th. M.~Nieuwenhuizen, V.~Spicka and
  A.~Yu. Khrennikov, editors, {\em Beyond the quantum}. World Scientific, 2007.

\bibitem{JCM}
E.~T. Jaynes and F.~W. Cummings.
\newblock {\em Proc. IEEE}, 51:89, 1963.

\bibitem{Scully1997}
Marlan~O. Scully and M.~Suhail Zubairy.
\newblock {\em Quantum Optics}.
\newblock Cambridge University Press, Cambridge, England, 1997.

\bibitem{Glauber1965}
R.~J. Glauber.
\newblock {\em Quantum Optics and Electronics, Proceedings of the Les Houches
  Summer School,}.
\newblock Gordon and Breach, New York, 1965.

\bibitem{Aquino2008}
Gerardo Aquino and Filippo Giraldi.
\newblock Entanglement entropy and determination of an unknown quantum state.
\newblock {\em Phys. Rev. A}, 78(6):062115, Dec 2008.

\bibitem{Shore1993}
B.~W. Shore and P.~L. Knight.
\newblock The jaynes-cummings model.
\newblock {\em ournal of Modern Optics}, 40(7):1195--1238, 1993.

\end{thebibliography}

%

\end{document}